\documentstyle[aps,twocolumn,graphicx]{revtex}

\begin{document}


\title{
Damping and frequency shift in the oscillations of two colliding 
Bose-Einstein condensates}

\author{M. Modugno, C. Fort, P. Maddaloni$^*$,
F. Minardi and M. Inguscio}

\address{INFM - European Laboratory for
  Non linear Spectroscopy (LENS) and Dipartimento di Fisica,\\
  Universit\`a di Firenze, Largo E. Fermi 2, I-50125 Firenze, Italy\\
$^*$Dipartimento di Fisica, Universit\`a di Padova,
Via F. Marzolo 8, I-35131 Padova, Italy}


\maketitle

\abstract{ We have investigated the center-of-mass oscillations of a
$^{87}$Rb Bose-Einstein condensate in an elongated magneto-static
trap.  We start from a trapped condensate and we transfer part of the
atoms to another trapped level, by applying a radio-frequency pulse.
The new condensate is produced far from its equilibrium position in
the magnetic potential, and periodically collides with the parent
condensate. We discuss how both the damping and the frequency shift of
the oscillations are affected by the mutual interaction between the
two condensates, in a wide range of trapping frequencies.  The
experimental data are compared with the prediction of a mean-field
model.}

\pacs{03.75.Fi, 05.30.Jp, 32.80.Pj, 34.20.Cf}

\section{Introduction}

The issue of the interaction between Bose-Einstein condensates in different
internal states has deserved much attention in the recent
literature, both from the experimental
\cite{jila1,jila2,jila3,jila4,otago,prl} 
and theoretical point of view \cite{sinatra,sinatra2,williams}.
The study of these mixtures of quantum fluids includes several
important topics, as the characterization of the dynamical behaviour, 
the response in presence of an external coupling, 
the phase coherence and the superfluid properties of the system.

Recently we have demonstrated an experimental me\-thod for a sensitive 
investigation of the interaction between two condensates 
\cite{prl,fringes,ramsey}, 
where two condensates are made to collide after periods of spatially 
separated evolution and quantitative information is extracted from the 
resulting collective dynamics.

In this Paper we report on new measures performed on such system
in a wide range of trapping frequencies, which allow 
a systematic investigation of the effects of the collisions between
the two condensates.

The experimental setup follows the scheme reported in Ref. \cite{prl}.
We use a radio-frequency (rf) pulse to produce two
$^{87}$Rb condensates in the states $|F=2, m_f=2\rangle\equiv |2\rangle$ and
$|2,1\rangle \equiv |1\rangle$. Due to the different magnetic moments
and the effect of gravity, they are trapped in two potentials whose
minima are displaced along the vertical $y$ axis by a distance much
larger that the initial size of each condensate. As a consequence the
$|1\rangle$ condensate, initially created with the same 
density distribution and at the same  position of the $|2\rangle$
condensate, undergoes large center-of-mass oscillations.
Moreover, the two condensates periodically collide, and this
strongly affects the collective excitations of both the condensates
\cite{prl}.
In particular the center-of-mass oscillation of the $|1\rangle$
condensate is damped and its frequency is shifted
upwards, with respect to the non interacting case.

The Paper is organized as follows.  In Section \ref{sec:setup},
we describe the experimental setup. Then, in Section
\ref{sec:oscillations}, we present the experimental results, 
analyzing the effect of mutual interactions 
on the center-of-mass motion over a wide range of trap frequencies.
The data are compared with the predictions of the Gross-Pitaevskii
theory for two coupled condensates at zero temperature, according to
the model discussed in Ref. \cite{pra}.
Finally, in Section \ref{sec:conclusion} we draw the conclusion.

\section{Experimental setup}
\label{sec:setup}

We start with a single $|2\rangle$ condensate containing typically 
$1.5\times 10^5$ $^{87}$Rb atoms at a temperature below 130~nK, 
confined in a 4-coils Ioffe-Pritchard trap elongated along the horizontal 
$z$ symmetry axis \cite{epj}, which produces an axially symmetric harmonic 
potential. The harmonic oscillator axial frequency is held fixed to 
$\nu_{z2}=12.6$~Hz, while the radial frequency $\nu_{\perp2}$ 
has been varied in the range $113\div310$ Hz by changing 
the minimum magnetic field $B_0$ according to the relation 
$\nu_{\perp2}=221\mbox{ Hz}/B_0 \mbox{[G]}^{1/2}$.  
The density distribution is an inverted parabola of the 
Thomas-Fermi (TF) regime \cite{bec_review}, and the typical radial and
longitudinal radii of the condensate are $R_\perp=4.3\div2.4~\mu$m
and $R_z=39\div58~\mu$m, in the range of frequencies considered here.

To populate the $|1\rangle$ state we apply a rf oscillating 
magnetic field. After the rf pulse the initial $|2\rangle$
condensate is put into a coherent superposition of different Zeeman
$|m_f\rangle$ sub-levels of the $F=2$ state, which then move apart: $|2
\rangle$ and $|1 \rangle$ are low-field seeking states and stay
trapped, $|0 \rangle$ is untrapped and falls freely under gravity,
while $|-1 \rangle$ and $|-2 \rangle$ are high-field seeking states
repelled from the trap.

By fixing the duration and varying the amplitude of the rf field, 
we control the relative population in different Zeeman sub-levels \cite{Erice}.  
In the experiments described here we use a 10 cycles rf
pulse at 1.24~MHz with an amplitude B=20~mG to quickly transfer
$\sim 13$\% of the atoms to the $|1\rangle$ state without populating
the $|0\rangle$, $|-1 \rangle$ and $|-2 \rangle$ states. 

In the frequency domain, the pulse width exceeds by almost one order of 
magnitude the transition broadening due to the inhomogeneous magnetic 
field across the condensate. As a result, the coupling strength does 
not depend on position and $|1\rangle$ is produced with the same spatial 
density profile as $|2\rangle$. Furthermore, since the characteristic time 
scale of the collective dynamics is of the order of 1~ms, no evolution 
occurs for the wavefunctions during the pulse, but, immediately after, 
excitations start in both the condensates.  

The $|1\rangle$ condensate experiences a trapping potential with lower
axial and radial frequencies ($\nu_1=\nu_2 / \sqrt{2}$), whose minimum
is displaced along the vertical $y$ axis by a distance 
$y_0 = g/ \omega_{\perp 2} ^2$ due to gravity 
(``gravitational sagging'').  In the frequency range considered here $y_0$
varies between 19 and 2.6~$\mu$m.
After the rf-pulse, the $|1\rangle$ condensate
moves apart from $|2\rangle$, and begins to oscillate around its
equilibrium position. Due to the mutual repulsive interaction, the
latter starts oscillating too, though with a much smaller amplitude.
Actually, each condensate moves in an effective potential which is the 
sum of the external potential (magnetic and gravitational) and the
mean-field one.  Here we have studied the dynamics of the $|1\rangle$ 
condensate by varying the permanence time in the trap up to 40~ms. 

After the release from the magnetic trap, we let 
the clouds drop and expand for 30~ms, and finally 
we take an absorption image on a CCD array. 
We notice that the process of switching off the magnetic field is not 
instantaneous, and lasts about 1 ms. Moreover it produces a magnetic gradient
which affects the initial velocity of the two condensates. 
The acquired velocities experimentally observed are 
$v_{1y}=0.7\pm0.1~$cm/s and $v_{2y}=1.4\pm0.1~$cm/s, 
respectively for the $|1\rangle$ and $|2\rangle$ condensates. 
In the following we use an unknown, effective, switch-off delay time as 
the only fitting parameter for our theoretical model.

\begin{figure}[t]
\centerline{\includegraphics[width=8cm]{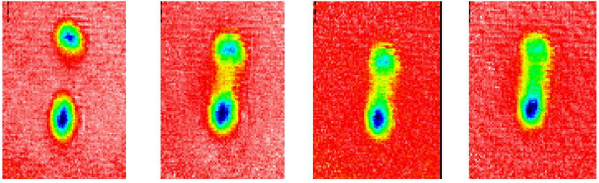}}
\caption{Images of the two condensates before and after the collision
(from left to right).}
\label{fig:twocond}
\end{figure}

\section{Center-of-mass oscillations}
\label{sec:oscillations}

Let us recall the basic features of our system, as reported
for the original experiment in Ref. \cite{prl}. 
The initial configuration corresponds to the stationary
ground-state with all the $N$ trapped atoms in
the $|2\rangle$ condensate.  Afterwards, at $t=0$, $N_1\simeq0.13 N$ atoms are
transferred from the $|2\rangle$ to the $|1\rangle$
state, the former remaining with $N_2=N-N_1$ atoms.

The dynamics of the condensate $|2\rangle$ is characterized by small amplitude
center-of-mass oscillations and by shape oscillations which are both
affected by the mutual repulsion with the other condensate when they 
periodically collide (see Fig. \ref{fig:twocond}) \cite{prl}. 

\begin{figure}[t]
\centerline{\includegraphics[height=8cm,angle=-90,clip=]{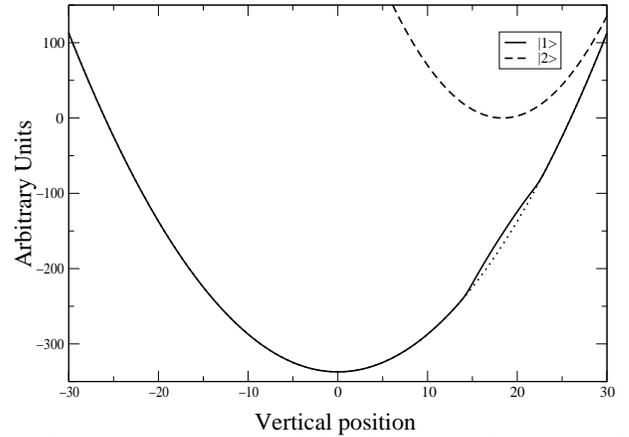}}
\caption{Trapping potentials for the two condensates in the vertical
direction for $\nu_{\perp2}=116$ Hz (lengths are in units of 
$a_{\perp2}= [\hbar/(m\omega_{\perp2})]^{1/2}\simeq1.00~\mu$m).
The $|1\rangle$ condensate moves in an effective potential (solid line)
which is modified by the mean field repulsion of $|2\rangle$ (right side).
}
\label{fig:pot}
\end{figure}

Here we are mainly interested in 
the motion of the $|1\rangle$ condensate, which is on the contrary 
dominated by large oscillations, whose frequency $\nu_{1}$ turns out to be 
larger than the ``bare'' trapping frequency $\nu_{\perp1}$. 
In fact the condensate $|1\rangle$ moves in an anharmonic effective potential
(see Fig.~\ref{fig:pot}) due to the presence of the condensate
$|2\rangle$, which provides an extra repulsion with a consequent
up-shift of frequency. 
This shift is one of the clear signatures of the mutual interaction,
since it only occurs if the two condensates overlap periodically. 

Furthermore, the oscillations appear to be damped. 
At the level of the Gross-Pitaevskii theory this damping is not 
due to dissipative processes (the total energy is conserved), but to
a redistribution of the kinetic energy initially associated 
with the center-of-mass motion of the $|1\rangle$ condensate, which is eventually 
shared among the other degrees of freedom of the system (center-of-mass of the other
condensate and internal excited states of both of them) \cite{pra}.

To investigate the role of the interactions in the large
center-of-mass oscillations of the $|1\rangle$ condensate in 
different dynamical regimes, we have carried out a systematic analysis by 
varying the trap frequency $\nu_{\perp1}$ in the range $80\div220$ Hz
(which corresponds to the above mentioned range $113\div310$ Hz for 
$\nu_{\perp2}$). 

\begin{figure}
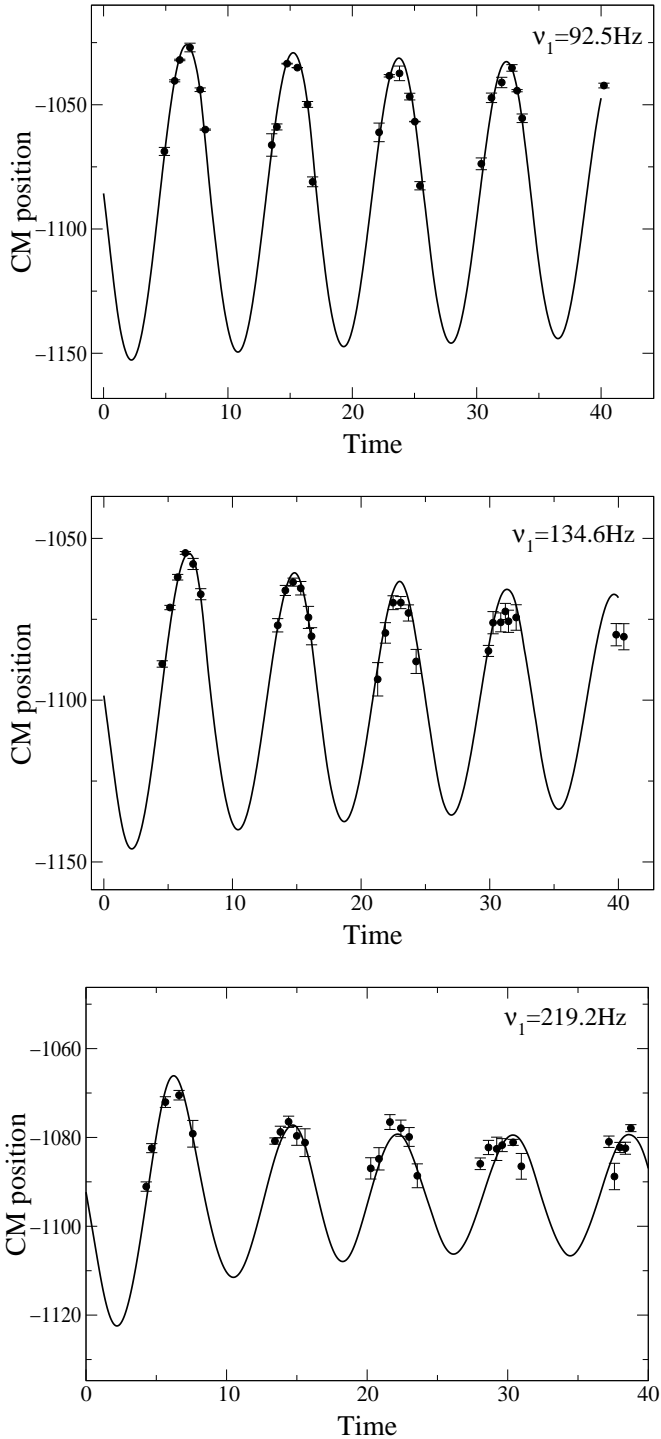

\centerline{\includegraphics[width=6.cm,angle=-90,clip=]{92.eps}}
\vspace{0.5cm}
\centerline{\includegraphics[width=6.cm,angle=-90,clip=]{134.eps}}
\vspace{0.5cm}
\centerline{\includegraphics[width=6.cm,angle=-90,clip=]{219.eps}}
\vspace{0.5cm}
\caption{
Center-of-mass oscillations of the $|1\rangle$ condensates 
after the ballistic expansion
for three values of the ``bare'' trapping frequency 
($\nu_{\perp1} = 92.1,134.6,219.2$ Hz,  from top to bottom). 
The measured center-of-mass positions (arbitrary units) are plotted as a 
function of the trapped evolution time (in units of $\omega_{\perp2}^{-1}$), 
and compared with the prediction of our 2D model (solid line).}
\label{fig:systematics}
\end{figure}

In Fig. \ref{fig:systematics} we show the center-of-mass motion of the $|1\rangle$
condensate, for some values of the
radial trapping frequency, by comparing the experimental data
with the prediction of the 2D Gross-Pitaevskii model discussed in
Refs. \cite{prl,pra}. In this case the only fitting parameter 
in the theoretical curves is the delay time caused by the non
instantaneous switch-off of the trapping potential. 
Its value is $\simeq 0.7$~ms. We notice also that in our model the
effect of the expansion on the center-of-mass motion is considered as  a
simple free fall under gravity, neglecting any residual interaction
between the condensates. Actually, although the two clouds acquire different 
velocities during the release from the trap, and cross each other in the fall, 
we expect them to quickly become so dilute, due to the fast radial expansion, 
that we can neglect their mutual mean-field repulsion. This is also supported by
numerical simulation including interaction in the early
stages of expansion \cite{nist:private}.

\begin{figure}
\centerline{\includegraphics[height=8cm,angle=-90,clip=]{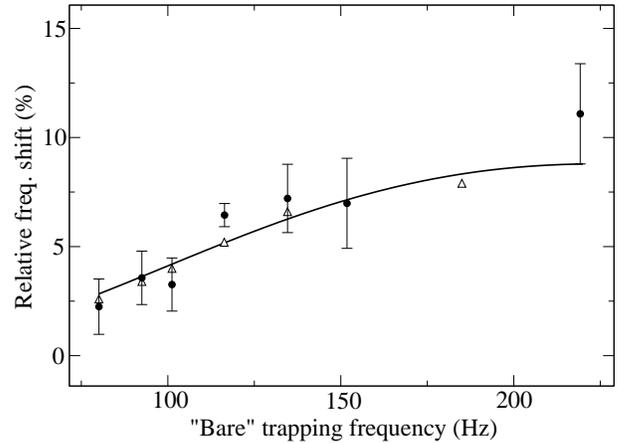}}
\vspace{0.5cm}
\caption{Relative shift of the $|1\rangle$ oscillation frequency versus 
  its ``bare'' trapping frequency, i.e. in absence of interactions 
  with $|2\rangle$. The experimental data are compared with the prediction 
of the classical  model (solid line) in Ref. \protect\cite{pra} and of
the 2D-GP model (empty triangles).}
\label{fig:fshift}
\end{figure}
 
\begin{figure}
\centerline{\includegraphics[height=8cm,angle=-90,clip=]{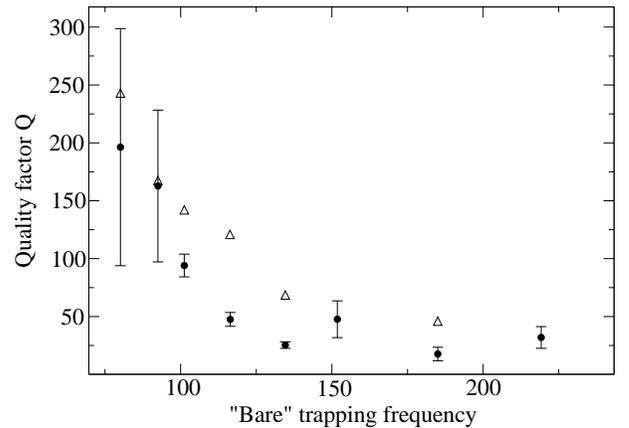}}
\vspace{0.5cm}
\caption{Measured ``quality factor'' $Q\equiv \omega_{\perp1}\tau$ 
for the $|1\rangle$ oscillation versus its ``bare'' frequency compared 
with the prediction of the 2D-GP model (empty triangles).} 
\label{fig:qfact}
\end{figure}

The values for the frequency shift and the damping time in the whole range 
of frequency are summarized in Figs. \ref{fig:fshift} and \ref{fig:qfact} 
respectively.  
In Fig. \ref{fig:fshift} the relative shift of the $|1\rangle$ oscillation 
frequency is plotted versus its ``bare'' trapping frequency, 
i.e. in absence of interactions with $|2\rangle$. The experimental data 
are compared with the prediction of the classical model (solid line) and
of the 2D-GP model (triangles).
In Fig.~\ref{fig:qfact}, we show the ``quality factor'' 
$Q=\omega_{\perp1}\cdot\tau$ of the oscillator as a function of  
$\nu_{\perp1}$, $\tau$ being the damping time, comparing
experimental points and theoretical predictions for our 2D model.

We notice that, concerning the frequency shift, there is a general
agreement (within the experimental uncertainty) between GP theory 
and experiment in the whole range of frequency, except for 
the frequency of the original experiment (116.3 Hz).
By the way, we have found that in this frequency range around 116.3 Hz 
there is an almost degeneracy between the energies of the two condensates
(mean field + kinetic + potential energy that the two condensates can
exchange during the collisions, see Ref. \cite{pra}). At this
stage, we cannot tell whether this can play some physical role, or it is
a pure coincidence.

On the contrary the observed damping is generally underestimated by
the 2D GP model. This is not surprising, since there could be other
sources of damping not included in our mean-field description.
To be more specific, one should consider the following points. 

{\em (i)}
The theoretical curves are based on a 2D model which only includes the 
radial degrees of freedom in the $xy$-plane. In fact, being the geometry of our
condensates strongly elongated in the longitudinal direction, we
have approximated the system as a uniform cylinder, by freezing
out the axial dynamics (see Ref. \cite{pra} for a comprehensive
discussion).
Therefore, the remaining contribution to the damping could in part arise
from the excitations of axial modes (possible distortions of the shape
of the two condensates along $z$), which are ignored in the
present analysis. 
Anyway, there are indications from preliminary 3D simulations that these
contributions should not be so important, at least for the case at
116.3 Hz.

{\em (ii)}
We observe a certain fragmentation of the $|1\rangle$ condensate after
some collision. This may have an influence on the measured
center-of-mass since the density distribution develops low density
asymmetric wings, which we might be unable to detect properly.  

{\em (iii)}
From the analysis of the experimental data there are indications in
support of the occurrence of atom (and energy) losses from both
condensates. In fact, the relative number of atoms for the 
$|1\rangle$ state ($N_1/N$) is in average a decreasing function of the
trapping time. We remind that, although each experimental point corresponds to a 
different experiment (each point corresponds to a condensates released from the 
trap after a variable time), they are all performed with almost the ``same'' 
initial conditions. In principle we can exclude that these losses
originate from the condensate finite lifetime ($0.7(1)$~s) or heating 
($dT/dt=0.11(2)~\mu$K/s), due to the short permanence times in the
trap (less than 40~ms).
Actually, this behaviour could be explained with 
a loss of atoms due to binary collisions between atoms of the two
condensates, an effect that is not included in our GP mean-field theory.

Concerning this last point, an appealing mechanism could be the
occurrence of elastic scattering processes between atoms of the two
overlapping condensates, as discussed in Ref. \cite{band} for the
impurities motion inside a Bose condensate \cite{mit:imp} and in the
analysis of matter 4-waves mixing \cite{nist:4wm}. In fact, since
Bose-Einstein condensates are superfluids, when the relative velocity
of the two condensates exceeds a critical value of the order of the
sound velocity, we could expect the GP mean-field approximation to
break down and ``elastic scattering'' losses to take place.  Indeed,
in our experiment, the two condensates collide at a velocity which is
larger than the (maximum) speed of sound at the top of $|2\rangle$,
$c_0=\hbar\sqrt{4\pi n_2 a_{22}}/m$, by a factor $\alpha$ that can be
tuned by changing the radial frequency. In the frequency range
considered, $\alpha$ goes from 4.2 at $\nu_{\perp2}=113$~Hz to 2.1 at
$\nu_{\perp2}=310$~Hz. So, we should expect the effect of the
``elastic scattering'', if any, to be more pronounced at low radial
frequency. As a consequence the results of the GP model should be more
accurate at high frequency. This conclusion, however, is not supported
by the experimental observations.

A detailed theoretical analysis of our experiment in this respect is
under way at the NIST group, in Gaithersbourg (USA). Preliminary
results give a better agreement for the damping time with the
inclusion of elastic scattering \cite{nist:private}, at least for the
case of the original experiment in Ref. \cite{pra}.

\section{Conclusions} 
\label{sec:conclusion}

To summarize, we have analyzed the large center-of-mass oscillations 
of a Bose-Einstein condensates in a strongly elongated magnetic trap,
over a wide range of frequencies. 
The condensate is produced by transferring part of the atoms from a
parent condensate in the $|F=2, m_f=2\rangle$ state to the
$|F=2, m_f=1\rangle$ state. The new condensate is created
far from its equilibrium position in the 
magnetic potential, and undergoes large oscillations, periodically
colliding with the parent condensate. 
We have shown that the mutual interactions
between the two condensates deeply affect both the amplitude and the
frequency of the oscillations. 
The experimental data are compared with the prediction of a 2D model based 
on the Gross-Pitaevskii theory for two coupled condensates at zero temperature.
This model accounts for the basic features of the system, and
the agreement with the data is generally satisfactory.  
Nevertheless, the discrepancy observed for the damping 
(and in some case also for the frequency) suggests that processes of
atom losses ({\em e.g.} elastic scattering losses \cite{band}) could 
play a significant role beyond the mean-field theory.
 
\section{Acknowledgments} 

We acknowledge useful discussions with Y. Band, J. Burke and 
P. Julienne. We also thank F. Dalfovo for critical readings of the manuscript.
This work was supported by the Cofinanziamento MURST and 
by the EU under Contract No. HPRICT1999-00111.
 

\end{document}